\documentclass[aip,apl,a4,reprint,twoside,floatfix,unsortedaddress]{revtex4-2}
\usepackage{graphicx}
\usepackage{color}
\usepackage{amsmath}
\usepackage{amsfonts}

\begin{document}

\title{Superconducting Properties of La$_{1.85}$Sr$_{0.15}$CuO$_4$ - Sr$_2$IrO$_4$ multilayers}
\author{Himanshu Pandey$^{1*}$, Manoj Kumar$^2$, D. Tripathi$^2$, and S. Pandey$^1$}

\affiliation{$^1$Department of Physics, Sardar Vallabhbhai National Institute of Technology, Surat-395007, India
\email{hp@phy.svnit.ac.in}
\\$^2$Department of Physics and Materials Science and Engineering, Jaypee Institute of Information Technology, Noida-201309, India}

\date{\today}

\begin{abstract}

We study superconducting properties in multilayer thin films consisting of superconducting La$_{1.85}$Sr$_{0.15}$CuO$_4$ (LSCO) and Mott insulator Sr$_2$IrO$_4$ (SIO) and report enhanced superconductivity in optimized sample. These multilayer heterostructures show an increase in superconducting transition temperature ($T_C$) as compared to the single layer LSCO films. The temperature dependence of SIO single layer is also investigated under thermal activation, Arrhenius-type behaviour, and variable-range hopping mechanisms for different temperature regimes. The decrease in $T_C$ beyond an optimum thickness of LSCO in these multilayers is analyzed in the framework of a model based on the assumption of induced superconductivity in SIO-LSCO interface due to the doping of La and/or oxygen deficiencies into SIO layers.

\end{abstract}

\maketitle

\section {Introduction}
The studies of transition metal oxide interfaces in the recent decade have led to the discovery of new two-dimensional electronic and magnetic phases stabilized by the broken inversion symmetry at the interface, direct and oxygen mediated overlap of transition metal orbits across the interface, charge transfer and even disorder and atomic intermixing.\cite{Ohtomo,Mannhart,PKR,Tsukazaki,Brinkman,Gozar,Seguchi,Fogel} In the context of superconductivity, the interface between two non-superconducting parent compounds can become a unique two-dimensional superconductor.\cite{Ohtomo,Mannhart,Brinkman,Reyren} While the origin of such behaviour is not fully understood, the observation gives strong impetus to research on new class of interfaces which may possibly sustain such behaviour. One transition metal oxide which has been catching attention from the perspective of stabilizing a superconducting phase is Sr$_2$IrO$_4$ (SIO).\cite{Moon,Yang} This 5$d$ transition metal oxide is structurally akin to the Mott insulator La$_2$CuO$_4$ which becomes superconducting on doping with Sr and Ba. The SIO is an antiferromagnet below $T_N$ = 240 K. A Mott insulating behaviour emerges from the strong spin - orbit interaction splitting of the $t_2g$ orbital into two groups of states labelled as J$_{eff}$ =1/2 and 3/2.\cite{Kim1,Kim2} Many similarities between La$_2$CuO$_4$ and SIO have led to theoretical works suggesting superconductivity on appropriate doping or structure modifications.\cite{YKKim,Chikara,Laguna,Cao1,Korneta,Wang,Shitade,Pesin,Lado} Indeed, some recent experiments on doped-SIO give strong indication of unusual metallic state,\cite{Torre} insulator-metal transition,\cite{Lee} resemblance with hole-doped cuprate superconductors,\cite{Yan} and incipient superconductivity in this compound.\cite{Nelson,Ilakovac,Brouet,Han}

Here, we describe a different strategy to induce superconductivity in SIO. Our experiment involves a layer-by-layer deposition of a superlattice of La$_{1.85}$Sr$_{0.15}$CuO$_4$ (LSCO) and SIO of the type [(LSCO)$_m$/(SIO)$_n$]$_5$ with fixed $n$ = 2 nm and $m$ varying from 2 nm to 50 nm. A significant increase in $T_C$ of $\approx$ 25 $\%$ is achieved  with respect to $m$ =2 and $\approx$ 10 $\%$ over the $T_C$ of bulk-like 100 nm thick LSCO film. We interpret these results in the framework of interfacial doping of the SIO.

\section {Experiment}

The growth of superconducting LSCO and Mott insulator SIO as well as their multilayer heterostructures was realized on (001) oriented SrLaO$_4$ (SLAO) substrates by using a multitarget pulsed laser deposition technique ($\lambda$= 248 nm) at 800$^\circ$C in 230 mTorr of oxygen pressure. After deposition of the samples, the deposition chamber was completely filled with oxygen to atmospheric pressure and then the samples were allowed to cool down to room temperature at a rate of 10$^\circ$C per minute with one-hour annealing at 500$^\circ$C to realize full oxygenation of the structure.\cite{PKR1} A film growth rate in the range of $\approx$ 0.006-0.012 nm/pulse was realized under these deposition conditions. The present study includes a series [(LSCO)$_m$/(SIO)$_n$]$_{\times5}$ heterostructures with $m$ = 2 nm to 50 nm while keeping the $n$ constant at 2 nm.

An X-ray diffractometer (PANalytical X'Pert PRO) equipped with Cu${K_{{\alpha _1}}}$ X-ray source with wavelength $\lambda$ = 0.154 nm, was used to study the crystallographic structure and interface quality of prepared multilayer samples in $\theta$-2$\theta$, $\omega$, $\varphi$, and X-ray reflectivity modes. The standard four-probe technique was used to measure in-plane superconducting response of the films through resistivity $\rho$(T) measurement using. Prior to transport measurement, Ar$^+$ ion milling process was utilized for sample patterning in four-probe geometry. The longitudinal voltage drop across the sample was measured for both positive and negative currents and then averaged out to correct for any thermal effects. For a good electrical contact and homogeneous current distribution across the ends of the thin film, silver contact pads were used. The current-voltage (I-V) response was also recorded at several temperatures in the presence of a magnetic field applied in parallel and also in perpendicular geometry with respect to the sample plane.

\section {Results and Discussion}

\begin{figure}[ht]
\begin{center}
\includegraphics [width=0.45\textwidth]{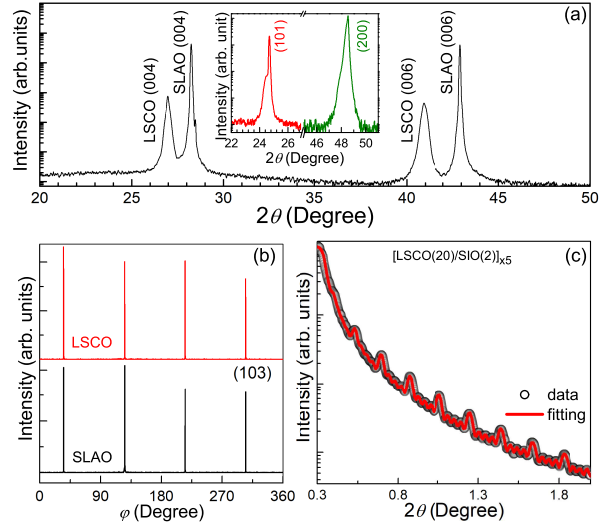}
\end{center}
\caption{\label{Fig 1} (Color online) (a) The $\theta-2\theta$ X-ray diffraction profile of [LSCO (10)/SIO (2)]$_{\times5}$ multilayer grown on SLAO in the vicinity of (006) and (008) reflections of the SLAO (001) substrate. The main intense peak is from the SLAO substrate whereas the left side shoulder peak is from the LSCO layer. The inset shows the X-ray diffraction profile measured about (101) and (200) peaks. (b) The $\varphi$-scans about (103) plane of LSCO layer and SLAO substrate. (c) The GIXR curve for [LSCO (20)/SIO (2)]$_{\times5}$ multilayer along with the fit. (Numbers in parentheses indicate respective layer thickness in nm).}
\end{figure}

The $\theta$-2$\theta$ X-ray diffraction profiles clearly reveal the presence of LSCO layer in the multilayer [LSCO(10 nm)/SIO(2 nm)]$_{\times5}$ as shown in Fig.~\ref{Fig 1}(a). The well-defined (006) and (008) Bragg peaks from the film are visible very close to the substrate peaks, indicating the (00$\ell$) oriented $c$-axis growth of LSCO layers. Rocking curves ($\omega$-scan) for the LSCO (006) reflection were measured to determine the out-of-plane mosaic spread. The values of full width at half maximum are less than 1.8$^\circ$ (data not shown here), suggesting that the samples have good crystallinity and strong (00$\ell$) texturing. The inset in Fig.~\ref{Fig 1}(a) shows the $\theta$-2$\theta$ scans about (101) and (200) peaks. From these scans the lattice parameter of LSCO layer are determined which are $a$ = 0.3771 nm, $b$ = 0.3778 nm, and $c$ = 1.3208 nm. Note that, there is a small biaxial compressive strain ($\epsilon_{xx}$ = $\epsilon_{yy}$ = -0.34$\%$) in the $ab$-plane resulting elongated $c$-axis of the LSCO layer. Further, the epitaxial growth of the multilayers has been confirmed by $\varphi$-scans. Figure~\ref{Fig 1}(b) shows the $\varphi$-scans measured about (103) peaks of LSCO film and SLAO substrate revealing sharp peaks only at integral multiple of $\pi/2$, suggesting the presence of four-fold cubic symmetry in these multilayer structures. The matching of the peaks for LSCO and SLAO in $\varphi$-scans indicates in-plane lattice matching with the epitaxial relation: [100] LSCO $||$ [100] SLAO. Though, we could not resolve the diffraction peaks from SIO the layer due to very thin layer of SIO used here, good resolved low-angle Kiessig fringes can be seen in the X-ray reflectivity curve for the sample [LSCO(20 nm)/SIO(2 nm)]$_{\times5}$ as shown in Fig.~\ref{Fig 1}(c), which not only provides the information about the uniform growth of LSCO-SIO superlattice structure but also used to estimate the thickness and roughness of the individual layers. The fitting of reflectivity curve yields the interface roughness between SLAO substrate and LSCO layer is of (0.3 $\pm$ 0.1) nm whereas between LSCO and SIO layers is $\approx$ (0.5 $\pm$ 0.1) nm. The measured layer thicknesses are in accord with thickness as estimated from the growth rate.

\begin{figure}[h]
\begin{center}
\includegraphics [width=0.45\textwidth]{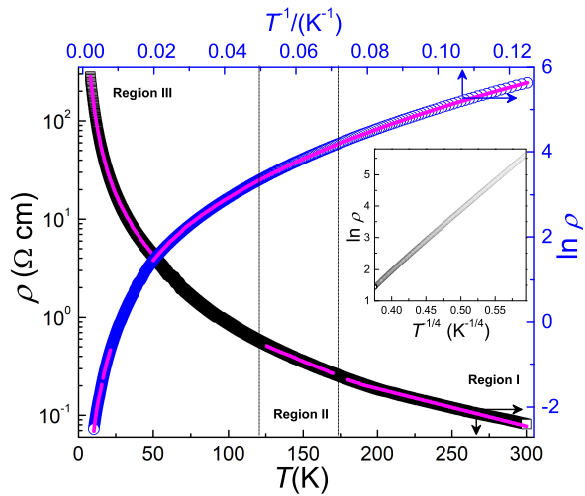}
\end{center}
\caption{\label{Fig 2} (Color online) Temperature dependence of resistivity for 10 nm thick SIO film. For the sake of clarity, $\ln \rho$ versus $T^{-1}$ is also plotted. The whole temperature range is divided in three different regions I, II and III. In the respective temperature ranges, the resistivity data is fitted with Eq.~\ref{rho-T}(a), (b) and (c), respectively (For more details see text). For better clarity of VRH fitting, a linear plot of $\ln \rho$ versus $T^{-1/4}$ is given in the inset.}
\end{figure}

Before moving towards the electrical transport investigation of these multilayers, we first present the temperature dependence of in-plane resistivity [$\rho(T)$] for 10 nm thick SIO single layer grown on SLAO substrate under the same deposition conditions as used for preparing the multilayers. Figure~\ref{Fig 2} shows the typical semiconducting behaviour ($d\rho/dT < 0$) of resistivity with a room-temperature resistivity is about 0.08 $\Omega$-cm. The similar order of magnitude of room temperature resistivity of single crystalline thin SIO film is reported.\cite{Nichols} The resistivity increases by three order of magnitude as the temperature is lowered to 5 K. We also plotted $\ln\rho$ versus $T^{-1}$ in the same fig.~\ref{Fig 2} for better understanding of $\rho(T)$. It was really very hard to fit the data in the whole range with single known model.\cite{Kini} We have identified three different temperature regions in which distinct temperature dependency were figured out. 1) 180 K $< T <$300 K (region-I), in which a logarithmic dependence [$\rho \propto$ exp(-$\alpha T$)] is observed; 2) 120 K $< T <$170 K (region-II), an Arrhenius-type behaviour, and 3) 20 K $< T <$50 K (region-III), a 3-\emph{d} variable-range-hopping (VRH) is observed. The governing equations for all these three regions are summarized below.

\begin{subequations}
\begin{align}
\rho \left( T \right) &= {\rho _0}\exp \left( { - \alpha T} \right) \qquad   180 K<T<300K\\
\rho \left( T \right) &= {\rho _0}\exp \left( {\frac{\Delta }{{{k_\beta }T}}} \right) \quad   120 K<T<170K\\
\rho \left( T \right) &= {\rho _0}\exp {\left( {\frac{{{T_0}}}{T}} \right)^{1/4}}  \quad  20 K<T<50K
\end{align}\label{rho-T} \end{subequations}

Here, $\rho_0$, $\alpha$, and $T_0$ are the fitting parameters; $k_{\beta}$ is the Boltzmann constant and $\Delta$ is the gap energy. For the region-I, the value of $\rho_0$ and $\alpha$ are 1.09 $\Omega$-cm and 8.81 $\times$ 10$^{-2}$ K$^{-1}$, respectively. The fitted data according to Eq.~\ref{rho-T}(a) is also plotted in Fig.~\ref{Fig 2}. The absence of Arrhenius-type behaviour in this temperature window, is confirmed by plotting $\ln \rho$ versus $T^{-1}$ data, which is not a straight line. Moving towards middle temperature range where a straight line fitting is obtained for $\ln \rho$ versus $T^{-1}$, indicating Arrhenius behaviour given by Eq.~\ref{rho-T}(b). The fitted data for this temperature range is also plotted in Fig.~\ref{Fig 2} and the fitting parameters are $\rho_0$ = 6.63 $\times$ 10$^{-2}$ $\Omega$-cm and $\Delta$ = 44 meV. A similar type of temperature dependence has been reported for epitaxial SIO thin films\cite{Nichols} as well as bulk crystal\cite{Ge} and our estimated value of gap energy is comparable with those reported values. It is worthful to mention here that the room temperature resistivity of 2 nm and 4 nm SIO single layer films was very high as compared to 10 nm SIO single layer film as for multilayer structures, 2 nm SIO was used in each deposition. In the low temperature range, the resistivity data is fitted with Eq.~\ref{rho-T}(c) and $\ln \rho$ versus $T^{-1/4}$ plot is found to be a straight line in this temperature range (see inset of Fig.~\ref{Fig 2}) which gives fitting parameters $\rho_0$ = 4.22 $\times$ 10$^{-3}$ $\Omega$cm and $T_0$ = 1.22 $\times$ 10$^{5}$ K. In this temperature region, this weaker temperature dependence suggests minimal long-range Coulomb repulsions between electrons. It is known that for VRH this exponent indicates the dimensionality of the conductor. The $T^{-1/(n+1)}$ dependence of electrical resistivity suggests an $n$-dimensional conductor. Hence, for our case, fitting shows a three-dimensional VRH conductivity. Cao \emph{et al.}\cite{Cao} also obtained the different VRH exponents 1/2 and 1/4 for a very wide temperature range for the case of single crystal of SIO.

\begin{figure}[t]
\begin{center}
\includegraphics [width=0.5\textwidth]{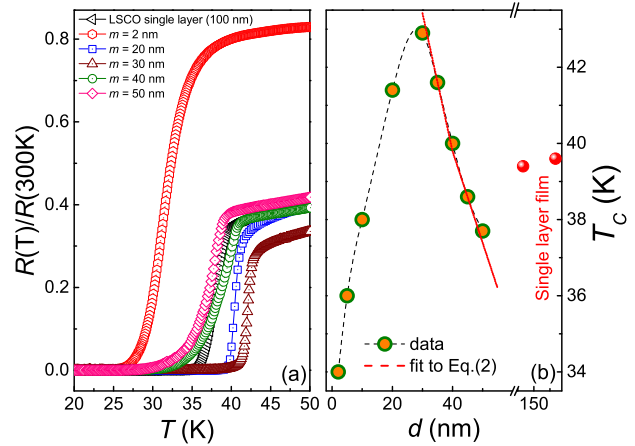}
\end{center}
\caption{\label{Fig 3} (Color online) (a) The temperature dependence of normalized resistance [\emph{R}(T)/\emph{R}(300K)]
for 100 nm thick single layer LSCO film and [LSCO (\emph{m})/SIO (2)]$_{\times5}$ multilayers with $m$ = 2, 20, 30, 40 and 50 nm, grown on SLAO substrate. (b) The variation of superconducting transition temperature ($T_C$) with LSCO layer thickness along with the $T_C$ values measured for 100 and 250 nm thick single layer LSCO film. Here, dashed line is only guide to eye and the full line is data fitting according to Eq. (3), as discussed in the text.}
\end{figure}

We now come to the central result of this paper as shown in Figure~\ref{Fig 3}(a) which depicts the temperature dependence of normalized resistance of different LSCO/SIO multilayers along with that of 100 nm thick single layer film of LSCO grown under the same deposition condition. The 100 nm LSCO film exhibits a $T_C$ onset around 39.4 K. We observe distinctly higher $T_C$ for the multilayer samples with LSCO thickness in the range $m$ = 20 to 40 nm as compared to that for single layer LSCO films of 100 and 250 nm thick. Here, the $T_C$ is defined as the onset temperature of the superconducting transition. The maximum $T_C$ enhancement of $\approx$ 43 K is observed for the sample with $m$ = 30 nm. A noteworthy increase in $T_C$ of $\approx$ 25 $\%$ is achieved with respect to the multilayer sample with $m$ = 2 nm and $\approx$ 10 $\%$ over the $T_C$ of bulk-like 100 nm thick LSCO film. The $T_C$ values increases with increasing LSCO film thickness and attain maximum (for $m$ = 30 nm) and then start to decrease as can be seen from Fig.~\ref{Fig 3}(b). Since, SIO layer is more resistive than LSCO layer, the $T_C$ in these multilayers will be sensitive to the sequence of the layers and also to the thickness of the each layers. This may suggest that the observed phenomenon is not related to interface. But there must be some perturbed interfacial region which may cause the variation in $T_C$ with LSCO layer thickness which will be discussed later.

\begin{figure}[h]
\begin{center}
\includegraphics [width=9 cm]{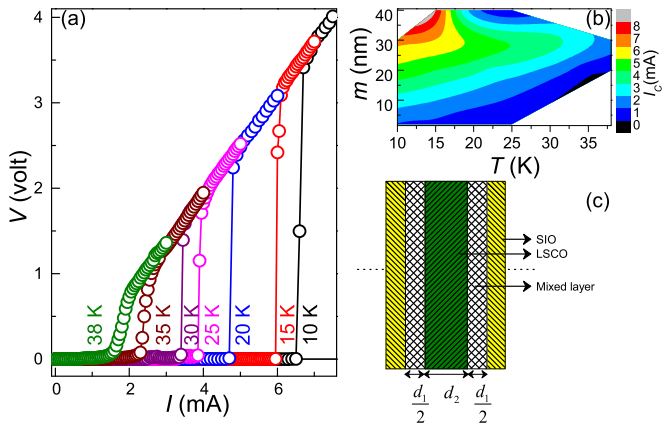}
\end{center}
\caption{\label{Fig 4} (Color online) (a) I-V characteristics for [LSCO(30)/SIO(2)]$_{\times5}$ multilayer measured at different temperature without applying any magnetic field. (b) The critical current distribution with temperature for multilayers with different LSCO layer thicknesses. (c) The schematic model for proximity effect between superconducting LSCO layer and mixed layer of SIO and LSCO. This mixed layer is assumed symmetrically on the both side of LSCO layer. (For more details, see text.) .}
\end{figure}

Figure~\ref{Fig 4}(a) shows isothermal I-V curves taken at different measurement temperatures for the sample with $m$ = 30 nm. We note that the superconducting critical current ($I_C$) depends on the thickness and the temperature of the sample. At a fixed value of $m$, $I_C$ reduces with increasing temperature and can be seen from the contour plot shown in Fig.~\ref{Fig 4}(b). This figure shows the observed variation of $I_C$ with temperature for different multilayer structures.

From Fig.~\ref{Fig 3}(a), we can see that the normal-state resistance of these multilayers changes with LSCO layer thickness suggesting some probability that the $T_C$ values are somehow affected by the proximity region within the individual superconducting LSCO layers. Hence, any kind of change or variation in normal-state properties of the LSCO layer will dominantly affect the superconducting properties of these multilayers. Apart form these, layer coupling due to minute doping from one layer to another or intermixing between the layers, may also be very relevant in our case. It would be very difficult to understand the mechanism of increase in $T_C$ via considering all aspects mentioned above.

To get into more depth about the variation in $T_C$, we have to look into the possibility of doping into SIO layer. As it is earlier proposed that the doping of electrons or holes in SIO can lead to achieve superconducting phases.\cite{Moon,Yang,Ilakovac} Since SIO is a $J = \frac{1}{2}$ Mott insulator with strong spin-orbit coupling, any minor change in stoichiometry results a notable change in the lattice parameters and also low-temperature transport properties due to doping and/or oxygen vacancies. Korneta \emph{et al.} reported the metallic behaviour on electron-doped (oxygen deficient) Sr$_2$IrO$_{4-\delta}$ with almost a nine order of huge magnitude change in the low temperature resistivity [$\rho$(T = 1.8 K)] as $\delta$ slightly changes from 0 to 0.04.\cite{Korneta} The substitution of La atoms on Sr-sites in SIO also makes it electron-doped which is also proposed to exhibit high-$T_C$ superconductivity.\cite{Wang}

From Fig.~\ref{Fig 3}(b), the decrease in $T_C$ of the multilayer system with increasing LSCO layer thickness beyond an optimum thickness of 30 nm, can be understood as a superconducting proximity effect. There are various calculations made for the proximity effect between thin layers. Cooper$-$de Gennes model\cite{Gennes} is one of them, which is based on the effective pairing interaction between the layers. But finite transparency of boundaries between different layers is not accounted for this model. So for the boundaries with smaller transparency, McMillan proposed a tunneling model\cite{McMillan} which further modified by Golubov\cite{Golubov} considering arbitrary transparency of the boundary via assuming the link between the different layers is single-particle tunneling. To apply the later model, one requires knowledge of $T_C$ and coherence length values for each layer, along with a parameter that governs the boundary conditions on the superconducting order parameter at the layers interface. Due to less knowledge of SIO layers as a superconducting candidate, it is really hard to use this model without knowing aforesaid values. We have tried to fit our experimental data using Gennes model and found a good agreement. To realize this, we have considered the possibility of intermixing of layers near the LSCO and SIO interface or doping/migration of La atoms into the SIO layer. Thus our multilayer samples can be understood in a sense where each of LSCO layer is covered both side by mixed LSCO-SIO layer and then entire this LSCO/mixed-layer structure is separated by insulating SIO layer. A sketch of this model is given in  Fig.~\ref{Fig 4}(c). These two regions with thicknesses $d_1$ and $d_2$, respectively, must have different values of $T_C$. This mixed layer will have different $T_C$ value other than its intrinsic value. The doping of La and/or oxygen deficiencies into SIO layers are the main basis of this model. To begin this, we need to know the effective pairing interaction in these two layers, which is given by
\begin{eqnarray}
{\lambda _{eff}} = \frac{{\sum\limits_{i = 1}^2 {{N_i}{d_i}{\lambda _i}} }}{{\sum\limits_{i = 1}^2 {{N_i}{d_i}} }}\;
\end{eqnarray}

where $N_i$, $d_i$, and $\lambda_i$ are the density of states at the Fermi level, the layer thickness and the pairing parameter, respectively, in the two layers. Here, layer 1 is the LSCO-SIO mixed layer and layer 2 is LSCO. In order to get correctly averaged pairing parameter, we assumed the presence of these mixed layers of same thicknesses on both sides of the LSCO layer. Thus $T_C$ of the system can be calculated from the formula given by\cite{McMillan}

\begin{eqnarray}
{T_C} = \frac{{{\Theta_D}}}{{1.45}}\exp \left[ {\frac{{ - 1.04\left( {1 + {\lambda_{eff}}} \right)}}{{{\lambda_{eff}} - {\mu^*}\left( {1 + 0.62{\lambda_{eff}}} \right)}}} \right]\;
\end{eqnarray}

with $\Theta_D$ is the Debye temperature and $\mu^*$ is the Coulomb coupling parameter. The fitted value of the parameter $N_1d_1/N_2$ is 6.22 nm, gives $d_1$ = 2.5 nm, which further indicates the presence of a $\approx$ 1.25 nm thick mixed layer on either side of the LSCO layer. Although this procedure is of course very approximate, the resulting values of $T_C$ for these multilayers systems with different LSCO layer thickness is in good agreement to the model and can be seen in Fig. ~\ref{Fig 3}(b). On this ground, we can say that with the decrease in LSCO layer thickness from 50 nm to 30 nm, Cooper pairs try to spend smaller time in LSCO layers and more in mixed layers due to proximity effect between the mixed and LSCO layer, this would increase the $T_C$ due to the dominance of larger pairing interaction in mixed layers. A similar kind of analysis was proposed and the superconducting transport properties were analyzed for Nb/Ge multilayer system by assuming passive influence on the superconducting order parameter associated with the proximity effect.\cite{Ruggiero} The presence of somewhat disordered Nb/Ge interfacial layer mainly controls the $T_C$ of Nb/Ge multilayers.

\section {Conclusions}
We have presented a detailed study of LSCO-SIO multilayers as a model system to understand the proximity effect between the layers. The temperature dependence of SIO single layer is also analyzed in the context of thermal activation, Arrhenius-type behaviour, and variable-range hopping mechanisms for different temperature regimes. With the increase in LSCO layer thickness, the LSCO-SIO multilayer system shows an increase in $T_C$ as high as $\approx$ 43K, 10 \% higher than the single layer LSCO film. After attaining a maximum, $T_C$ decreases with further increasing LSCO layer thickness. For smaller LSCO thickness Cooper pairs spends lesser time due to the dominance of larger pairing interaction in the mixed layers. Although we have tried to give the explanation for the thickness dependence of $T_C$, the poor information about SIO binds us from detailed investigation. A high resolution interface study is required for the better understanding. Our results definitely will give a direction to achieve proposed superconductivity in Sr$_2$IrO$_4$ and also excite the intriguing field of layer superconductors.

\section {Acknowledgements}
H.P. thanks to Mr. P. C. Joshi for his help in Ar$^+$-ion milling and to Prof. R. C. Budhani and Prof. Z. Hossain for valuable discussions and making available laboratory facilities. H.P. acknowledges Science and Engineering Research Board (SERB), Government of India for the research grant against early career research award scheme ECR/2017/001612.


\begin{thebibliography}{100}


\bibitem{Ohtomo}
A. Ohtomo and H. Y. Hwang, Nature (London) \textbf{427}, 423 (2004).

\bibitem{Mannhart}
J. Mannhart and D. G. Schlom, Science \textbf{327}, 1607 (2010).

\bibitem{PKR}
P. K. Rout, H. Pandey, L. Wu, Anupam, P. C. Joshi, Z. Hossain, Y. Zhu, and R. C. Budhani, Phys. Rev. B \textbf{89}, 020401(R) (2014).

\bibitem{Tsukazaki}
A. Tsukazaki, A. Ohtomo, A. T. Kita, Y. Ohno, H. Ohno, and M. Kawasaki, Science \textbf{315}, 1388 (2007).

\bibitem{Brinkman}
A. Brinkman, M. Huijben, M. van Zalk, J. Huijben, U. Zeitler, J. C. Maan, W. G. van der Wiel, G. Rijnders, D. H. A. Blank, and H. Hilgenkamp, Nat. Mater. \textbf{6}, 493 (2007).

\bibitem{Gozar}
A. Gozar, G. Logvenov, L. F. Kourkoutis, A. T. Bollinger, L. A. Giannuzzi, D. A. Muller, and I. Bozovic, Nature (London) \textbf{455}, 782 (2008).

\bibitem{Seguchi}
Y. Seguchi, T. Tsuboi, and T. Suzuki, J. Phys. Soc. Jpn. \textbf{61}, 1875 (1992).

\bibitem{Fogel}
N. Ya. Fogel, E. I. Buchstab, Yu. V. Bomze, O. I. Yuzephovich, M. Yu. Mikhailov, A. Yu. Sipatov, E. A. Pashitskii, R. I. Shekhter, and M. Jonson, Phys. Rev. B \textbf{73}, 161306(R) (2006).

\bibitem{Reyren}
N. Reyren, S. Thiel, A. D. Caviglia, L. F. Kourkoutis, G. Hammer, C. Richter, C. W. Schneider, T. Kopp, A.-S. Ruetschi, D. Jaccard, M. Gabay, D. A. Muller, J.-M. Triscone, and J. Mannhart, Science \textbf{317}, 1196 (2007).

\bibitem{Moon}
S. J. Moon, M. W. Kim, K. W. Kim, Y. S. Lee, J.-Y. Kim, J.-H. Park, B. J. Kim, S.-J. Oh, S. Nakatsuji, Y. Maeno, I. Nagai, S. I. Ikeda, G. Cao, and T. W. Noh, Phys. Rev. B \textbf{74}, 113104 (2006).

\bibitem{Yang}
Y. Yang, W.-S. Wang, J.-G. Liu, H. Chen, J.-H. Dai, and Q.-H. Wang, Phys. Rev. B \textbf{89}, 094518 (2014).

\bibitem{Kim1}
B. J. Kim, H. Jin, S. J. Moon, J.-Y. Kim, B.-G. Park, C. S. Leem, J. Yu, T. W. Noh, C. Kim, S.-J. Oh, J.-H. Park, V. Durairaj, G. Cao, and E. Rotenberg, Phys. Rev. Lett. \textbf{101}, 076402 (2008).

\bibitem{Kim2}
B. J. Kim, H. Ohsumi, T. Komesu, S. Sakai, T. Morita, H. Takagi, and T. Arima, Science \textbf{323}, 1329 (2009).

\bibitem{YKKim}
Y. K. Kim, O. Krupin, J. D. Denlinger, A. Bostwick, E. Rotenberg, Q. Zhao, J. F. Mitchell, J. W. Allen, and B. J. Kim, Science \textbf{345}, 187 (2014).

\bibitem{Chikara}
S. Chikara, O. Korneta, W. P. Crummett, L. E. DeLong, P. Schlottmann, and G. Cao, Phys. Rev. B \textbf{80}, 140407(R) (2009).

\bibitem{Laguna}
M. A. Laguna-Marco, D. Haskel, N. Souza-Neto, J. C. Lang, V. V. Krishnamurthy, S. Chikara, G. Cao, and M. van Veenendaal, Phys. Rev. Lett. \textbf{105}, 216407 (2010).

\bibitem{Cao1}
G. Cao, V. Durairaj, S. Chikara, L. E. DeLong, S. Parkin, and P. Schlottmann, Phys. Rev. B \textbf{76}, 100402(R) (2007).

\bibitem{Korneta}
O. B. Korneta, T. Qi, S. Chikara, S. Parkin, L. E. De Long, P. Schlottmann, and G. Cao, Phys. Rev. B \textbf{82}, 115117 (2010).

\bibitem{Wang}
F. Wang and T. Senthil, Phys. Rev. Lett. \textbf{106}, 136402 (2011).

\bibitem{Shitade}
A. Shitade, H. Katsura, J. Kune\v{s}, X.-L. Qi, S.-C. Zhang, and N. Nagaosa, Phys. Rev. Lett. \textbf{102}, 256403 (2009).

\bibitem{Pesin}
D. Pesin and L. Balents, Nat. Phys. \textbf{6}, 376 (2010).

\bibitem{Lado}
J. L. Lado and V. Pardo, Phys. Rev. B \textbf{92}, 155151 (2015).

\bibitem{Torre}
A. de la Torre, S. M. Walker, F. Y. Bruno, S. Ricc\'{o}, Z. Wang, I. G. Lezama, G. Scheerer, G. Giriat, D. Jaccard, C. Berthod, T. K. Kim, M. Hoesch, E. C. Hunter, R. S. Perry, A. Tamai, and F. Baumberger, Phys. Rev. Lett. \textbf{115}, 176402 (2015).

\bibitem{Lee}
J. S. Lee, Y. Krockenberger, K. S. Takahashi, M. Kawasaki, and Y. Tokura, Phys. Rev. B \textbf{85}, 035101 (2012).

\bibitem{Yan}
Y. J. Yan, M. Q. Ren, H. C. Xu, B. P. Xie, R. Tao, H. Y. Choi, N. Lee, Y. J. Choi, T. Zhang, and D. L. Feng, Phys. Rev. X \textbf{5}, 041018 (2015).

\bibitem{Nelson}
J. N. Nelson, C. T. Parzyck, B. D. Faeth, J. K. Kawasaki, D. G. Schlom, and K. M. Shen, Nat. Comm. \textbf{11}, 2597 (2020).

\bibitem{Ilakovac}
V. Ilakovac, A. Louat, A. Nicolaou, J.-P. Rueff, Y. Joly, and V. Brouet, Phys. Rev. B \textbf{99}, 035149 (2019).

\bibitem{Brouet}
V. Brouet, J. Mansart, L. Perfetti, C. Piovera, I. Vobornik, P. Le F\`{e}vre, F. Bertran, S. C. Riggs, M. C. Shapiro, P. Giraldo-Gallo, and I. R. Fisher

\bibitem{Han}
T. Han, D. Liang, Y. Wang, J. Yang, H. Han, J. Wang, J. Gong, L. Luo, W. K. Zhu, C. Zhang, and Y. Zhang, J. Supercond. Nov. Magn. \textbf{30}, 3493 (2017).

\bibitem{PKR1}
P. K. Rout, P. C. Joshi, R. Porwal1, and R. C. Budhani, Europhysics Letters \textbf{98}, 67007 (2012).

\bibitem{Kini}
N. S. Kini, A. M. Strydom, H. S. Jeevan, C. Geibel, and S. Ramakrishnan, J. Phy. Con. Mat. \textbf{18}, 8205 (2006).

\bibitem{Nichols}
J. Nichols, O. B. Korneta, J. Terzic, L. E. De Long, G. Cao, J. W. Brill, and S. S. A. Seo, App. Phys. Lett. \textbf{103} 131910 (2013).

\bibitem{Ge}
M. Ge, T. F. Qi, O. B. Korneta, D. E. De Long, P. Schlottmann, W. P. Crummett, and G. Cao, Phys. Rev. B \textbf{84}, 100402(R) (2011).

\bibitem{Cao}
G. Cao, J. Bolivar, S. McCall, J. E. Crow, and R. P. Guertin, Phys. Rev. B \textbf{57}, R11039 (1998).

\bibitem{Gennes}
P. G. de Gennes, Rev. Mod. Phys. \textbf{36}, 225 (1964).

\bibitem{McMillan}
W. L. McMillan, Phys. Rev. \textbf{175}, 542 (1968).

\bibitem{Golubov}
A. A. Golubov, Proc. SPIE 2157, Superconducting Superlattices and Multilayers, 353 (1994).

\bibitem{Ruggiero}
S. T. Ruggiero, T. W. Barbee, Jr. and M. R. Beasley, Phys. Rev. B \textbf{26}, 4894 (1982).

\end{thebibliography}
\end{document}